\documentclass{PoS}

\usepackage{color}
\usepackage{graphicx}
\usepackage{amsmath}
\usepackage{array}
\usepackage{subcaption}
\usepackage{float}
\usepackage[justification=raggedright]{caption}

\title{Search for dark matter from the first data of the PandaX-II experiment}
\ShortTitle{}

\author{\speaker{Yong Yang}\thanks{On behalf of PandaX-II Collaboration}\\
 INPAC and Department of Physics and Astronomy, Shanghai Jiao Tong University, Shanghai Laboratory for Particle Physics and Cosmology, Shanghai 200240, China \\
        E-mail: \email{yong.yang@sjtu.edu.cn}}

\abstract{Results of WIMP dark matter search from the first data of
the PandaX-II experiment are presented. PandaX-II experiment uses
a 500 kg scale dual phase liquid xenon time projection chamber,
operating at the China JinPing Underground Laboratory. The first data
correspond to a total exposure of $3.1\times10^{4}$ kg-day. The
observed data after selections are found to be consistent with
background expection, and upper limits of the spin-independent
WIMP-nucleon cross sections are derived for a range of WIMP mass
between 5 GeV/c$^2$ and 1000 GeV/$c^{2}$. The lowest cross section
limit obtained is 2.5$\times$10$^{-46}$~cm$^2$ at a WIMP mass of 40
GeV/c$^2$.}

\FullConference{38th International Conference on High Energy Physics\\
		3-10 August 2016\\
		Chicago, USA}

\begin{document}

\section{Motivation and Overview of PandaX-II}

The existence of dark matter (DM) in the Universe has been established
by overwhelming astronomical and cosmological evidences. However its
particle nature remains to be elusive. One of the most promising
candidates for DM is the weakly interacting massive particles (WIMPs),
a class of hypothetical particles predicted by many extensions of the
standard model of particle physics.  Detection for WIMP signals has
been the goal of many past, ongoing and future experiments, including
DM direct detection experiments, indirect detection experiments and
experiments at high energy particle colliders.

The PandaX project consists of a series of xenon-based
experiments. PandaX-I experiment, the first phase the project, uses a
120-kg liquid xenon target to search for WIMPs. It has completed in
2014 and produced stringent limits on the WIMP-nucleon cross sections
for low mass WIMPs~\cite{Xiao:2014xyn, Xiao:2015psa}.  PandaX-II, with
a half-ton scale xenon target, has started to take data since the end
of 2015.  The third experiment PandaX-III~\cite{Chen:2016qcd}, which
is being prepared, will search for neutrinoless double beta decay of
$^{136}$Xe.  Both PandaX-I and PandaX-II experiments use the
dual-phase xenon time projection chamber (TPC) technique to search for
WIMPs. This technique allows the measurement of both the prompt
scintillation photons (S1) produced in liquid xenon and the delayed
electroluminescence photons (S2) produced in gas xenon for each
physical event, leading to powerful background suppression and
signal-background discrimination.

The experimental hall for PandaX-I and PandaX-II experiments is
located at the China JinPing underground Laboratory (CJPL), which is
the deepest underground laboratory in the world, where the muon flux
is only about 60 events /m$^2$/year due to the above shield of 2400 m
of rocks or 6800 m water equivalent depth. The background due to
muon-induced neutrons for DM search is negligible.

PandaX-II reuses most of the infrastructures of PandaX-I, such as the
passive shielding, outer copper vessel, cryogenics and electronics and
data acquisition system. The new inner vessel is constructed
from stainless steel with much lower radioactivity, reducing the
$^{60}$Co activity by more than an order of magnitude. A new and
larger xenon TPC is also constructed. It contains 580~kg liquid xenon
in the sensitive volume enclosed by polytetrafluoroethylene (PTFE)
reflective panels with an inner diameter of 646~mm and a vertical
maximum drift length of 600~mm defined by the cathode mesh and gate
grid. For each physical event, both S1 and S2 signals are collected by
two arrays of 55 Hamamatsu R11410-20 photomultiplier tubes (PMTs)
located at the top and bottom, respectively.  A skin liquid xenon
region outside of the PTFE wall was instrumented with 48 Hamamatsu
R8520-406 1-inch PMTs serving as an active veto.  More detailed
descriptions of the PandaX-II experiments can be found in
Ref.~\cite{Tan:2016diz}.

\section{Data and Detector Calibration}
The data for DM search presented in this proceeding consist of 19.1
and 79.6 effective days of data collected during the first
commissioning run (Run 8, Nov. 21 to Dec. 14, 2015) and the first
physics run (Run 9, March 9 to June 30, 2016) of PandaX-II,
respectively. Taking into account the fiducial volume (FV, defined in next
section), this data set corresponds to a total exposure of
$3.1\times10^{4}$ kg-day.

\begin{figure}[h] \centering
  \includegraphics[width=.49\textwidth]{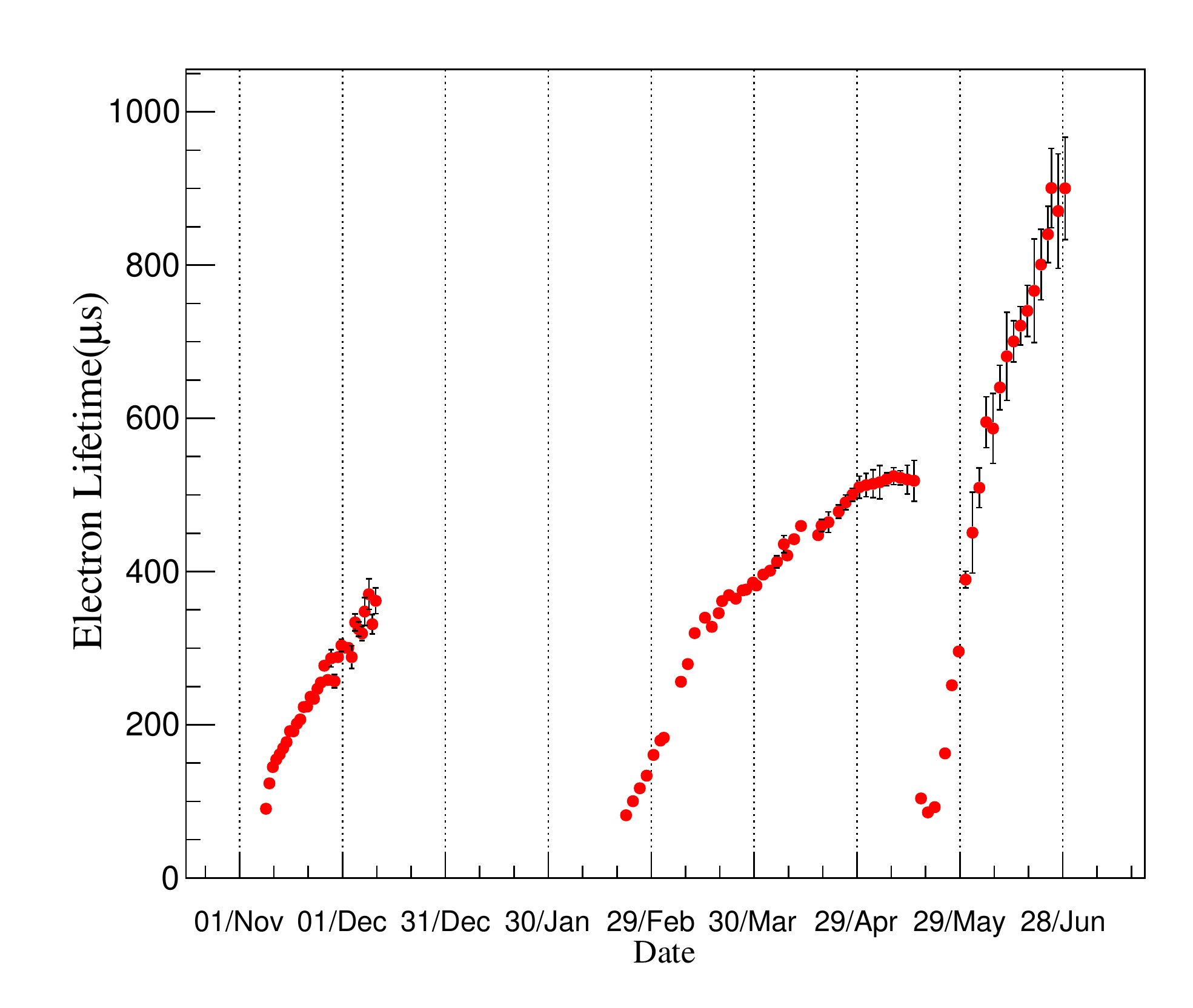}
  \includegraphics[width=.49\textwidth]{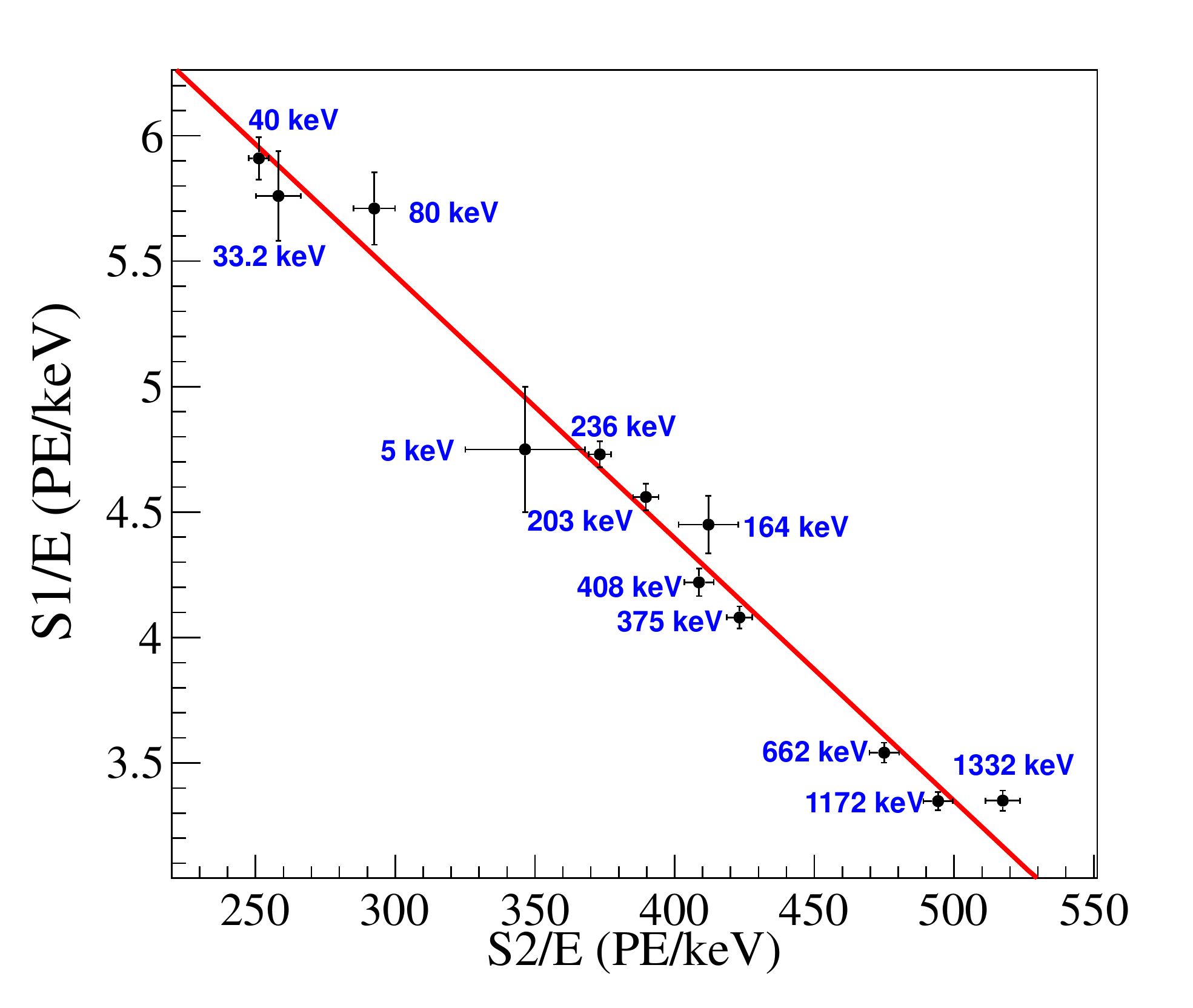}
  \caption{Left, evolution of the electron lifetime in Run 8 and Run
9. Right, Linear fit in S2$/E$ vs. S1$/E$ for all ER peaks in data to
determine the PDE and EEE.}
  \label{fig:e_lifetime}
\end{figure}

To calibrate the detector response, a neutron source ($^{252}$Cf) and
two $\gamma$ sources ($^{60}$Co and $^{137}$Cs) were deployed through
two PTFE tubes at different heights surrounding the inner
vessel. Neutrons can excite xenon nuclei or produce metastable nuclear
states, leading to de-exciting $\gamma$ rays at 40 ($^{129}$Xe), 80
($^{131}$Xe), 164 ($^{131m}$Xe), and 236 keV
($^{129m}$Xe). Photo-absorption $\gamma$ peaks were used to calibrate
the detector response. The 164~keV $\gamma$ events were uniformly
distributed in the detector and were used to produce a uniformity
correction for the S1 and S2 signals. A 3-D correction map was
produced for S1. For the S2 signals, the vertical uniformity
correction was obtained by fitting S2 vs. the drift time using an
exponential decay constant $\tau$, known as the electron lifetime,
which is a key parameter to indicate the purity of the liquid
xenon. The evolution of the measured electron lifetime in Run 8 and
Run 9 is shown in Fig.~\ref{fig:e_lifetime} left. Only data with
electron lifetime longer than 205 $\mu$s were used for the DM search.

After the above uniformity corrections for S1 and S2, the deposited
energy (electron equivalent energy $E_{ee}$) of each event is
reconstructed as
\begin{equation} E_{ee} = W\times\left(\frac{\text{S1}}{\text{PDE}} +
\frac{\text{S2}}{\text{EEE}\times\text{SEG}}\right)\,,
\end{equation} where $W = 13.7$~eV is the average work function to
produce either an electron or photon~\cite{Lenardo:2014cva}.
$\text{PDE}$, the photon-detection efficiency, $\text{EEE}$, the
electron extraction efficiency, and $\text{SEG}$, the single-electron
gain in PE/e, are the three key detector parameters to be
determined. $\text{SEG}$ can be determined from the distribution of
the smallest S2 signals in the data that were identified as the single
electron signals. Then the other two parameters PDE and EEE can be
extracted from fitting the S2$/E$ vs S1$/E$ obtained from all $\gamma$
peaks, shown in Fig.~\ref{fig:e_lifetime} right. Extraction of these
detector parameters was carried out for each detector running
condition, arising from different TPC field settings that were used to
maximize the drift and electron extraction fields while avoiding
spurious photons and electrons emission from the electrodes.

\begin{figure}[h] \centering
  \includegraphics[width=0.49\textwidth]{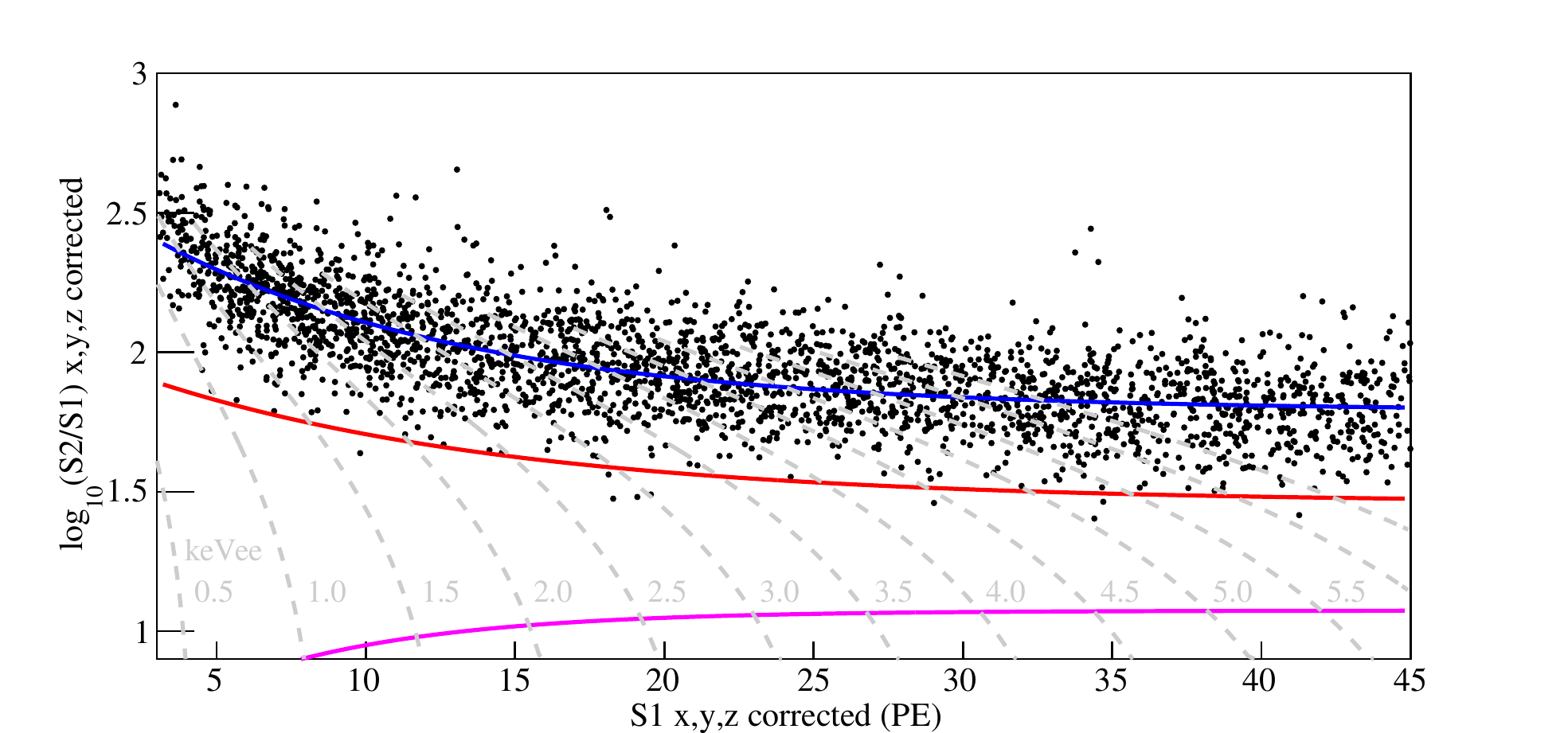}
  \includegraphics[width=0.49\textwidth]{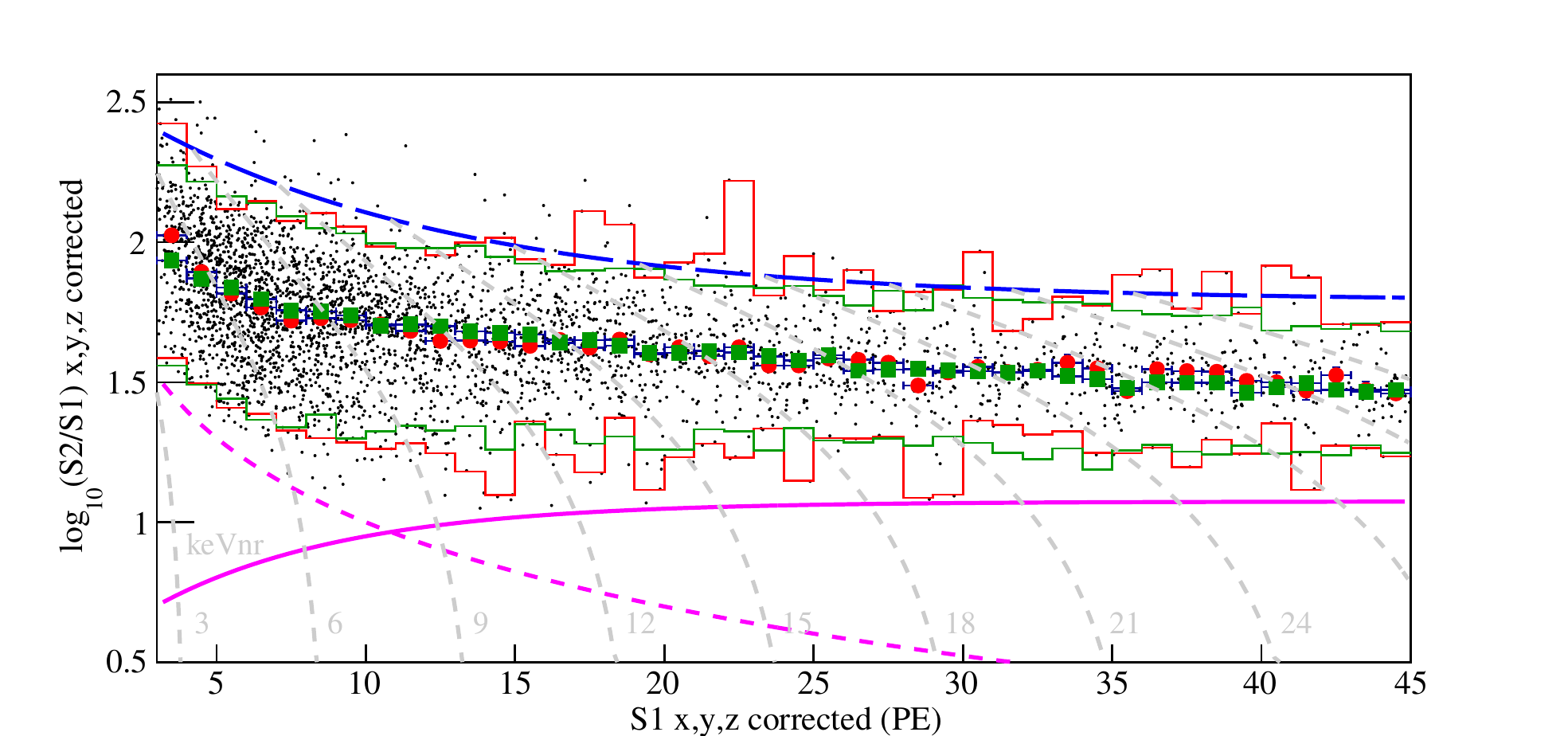}
  \caption{Left: tritium calibration data in $\log_{10}(\text{S2/S1})$
vs. S1, and fits of medians of ER (blue) and NR (red) data.  Right:
AmBe calibration data in $\log_{10}(\text{S2/S1})$ vs. S1, together
with medians from the data (red solid circles) and simulation (green
squares), and the fit to ER medians (blue dashed). The 2.3- and
97.7-percentiles from the data (red lines) and simulation (green
lines) are overlaid.}
\label{fig:calibration}
\end{figure}

In Run 9, nuclear recoil (NR) calibrations were performed using a
low-intensity (approximately 2 Hz) $^{241}$Am-Be (AmBe) neutron source
with improved statistics, and a low energy electron recoil (ER)
calibrations were performed by injecting tritiated
methane~\cite{Akerib:2015wdi}, leading to a better understanding of
the distributions of the NR and ER events than before.  Distributions
of $\log_{10}(\text{S2/S1})$ vs. S1 from these calibration data after
selections (described in next section) are shown in
Fig.~\ref{fig:calibration}, which also shows the medians and widths of
$\log_{10}(\text{S2/S1})$ from NR data and simulation. Reasonably good
agreement is observed.

\section{Event Selections and Background Estimation}
The following selections are applied. All collected events are
required to pass a set of data quality selections to remove noises and
abnormal events. Only events with single S2 are selected, and S2 is
required to be between 100 photoelectrons (PE) and 10000 PE. The
maximum number of S1 signals is limited to two, and the maximum one is
chosen to pair with S2. The S1 is required to be between 3 PE and 45
PE.  The event is vetoed if there is a coincidence between signals
from veto PMT arrays and the S1. The reconstructed event vertex is
required to be inside the FV. The vertex position in
horizontal plane is required to be less than 268 mm. The drift time is
required to be between 20 to 346 $\mu$s for Run 8, and between 18 to
310 $\mu$s for Run 9, where a more stringent maximum drift time cut is
needed to suppress the below-cathode $\gamma$ energy deposition
(so-called ``gamma-X'') from $^{127}$Xe decays which were not present
in Run 8.

Background of DM searches in PandaX-II can be separated into three
types: ER, neutron, and accidental background.  In the first data, the
ER background is dominated by decays of $^{85}$Kr which was likely
introduced by an air leak during the previous fill and recuperation
cycle before Run 8, and by decays of $^{127}$Xe which was generated
during the krypton distillation campaign in early 2016, when the xenon
was exposed to about one month of sea level cosmic ray
radiation. $^{85}$Kr background is identified using the delayed $\beta-\gamma$
coincidence from $^{85}$Kr decay. $^{127}$Xe background is identified by 
the 33 keV and 5.2 keV X rays originating from $^{127}$Xe decays. A parameterized tritium event distribution was used
to simulate expected distributions for different ER background
components. Neutron background comes from detector components and is
estimated from simulation. Accidental background is estimated by
randomly pairing single S1 and single S2 events. A multivariate
technique is developed to suppress this background. A factor of three
rejection is achieved while maintaining a 90\% signal efficiency.

\section{Results}

\begin{figure}[h] \centering
  \includegraphics[width=.49\textwidth]{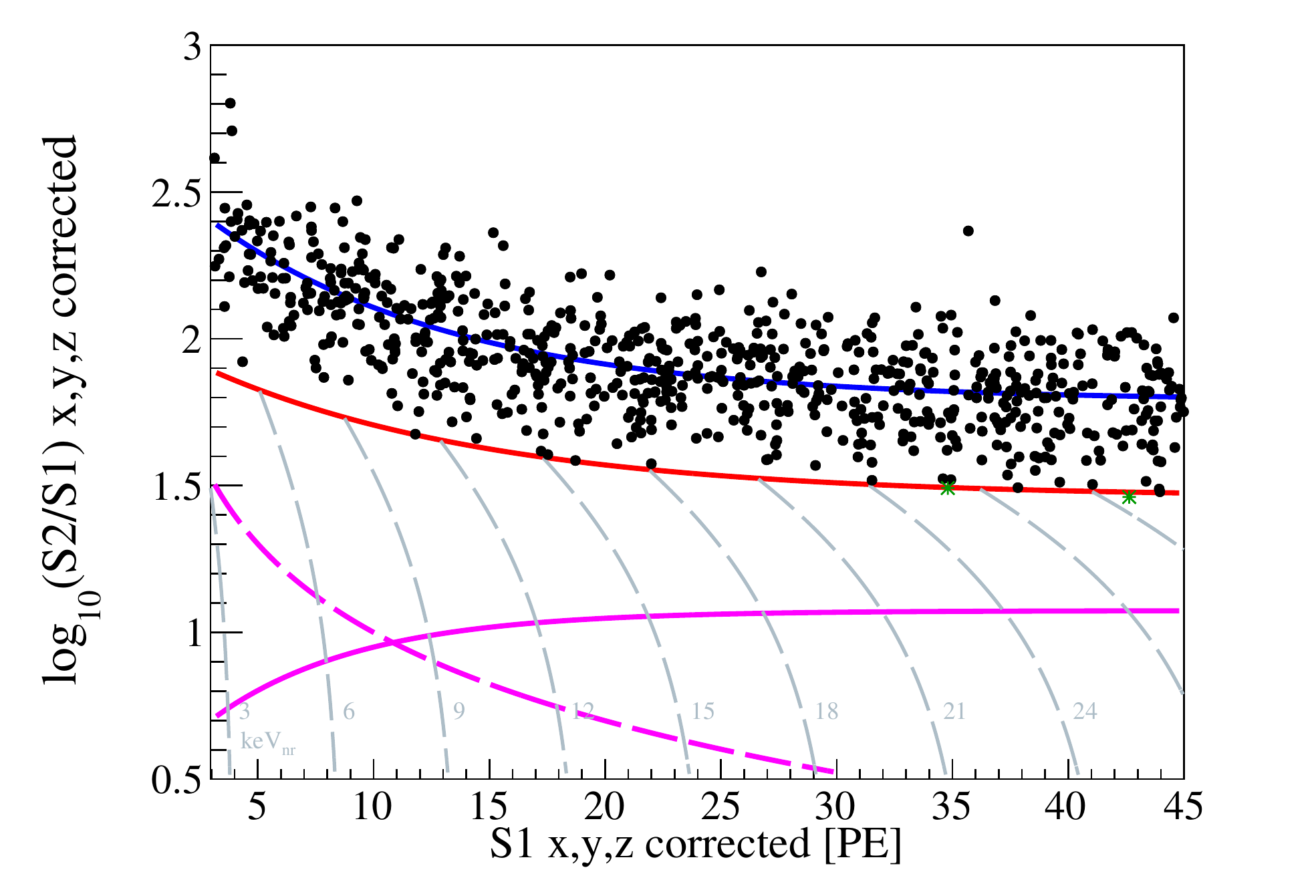}
  \includegraphics[width=0.49\textwidth]{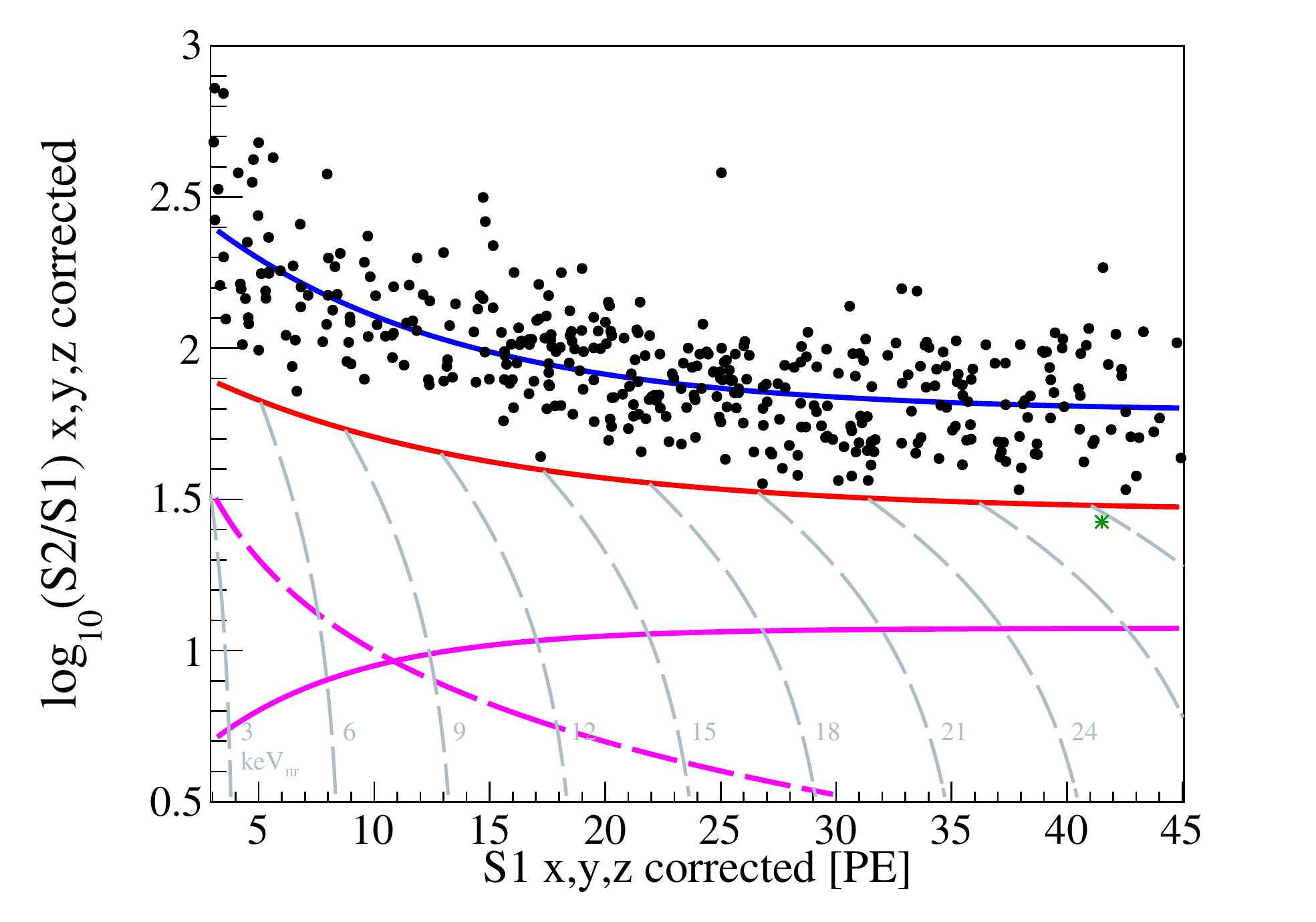}
  \caption{The distribution of $\log_{10}$(S2/S1) versus S1 for the DM
search data in Run 8 (left) and Run 9 (right).  The median of the NR
calibration band is indicated as the red curve. The dashed magenta
curve represents the equivalent 100~PE cut on S2. The solid magenta
curve is the 99.99\% NR acceptance curve.  The gray dashed curves
represent the equal energy curves with NR energy indicated in the
figures. Data points below the NR median curve are highlighted as a
green star.}
  \label{fig:dm_band}
\end{figure}

The observed events in data and expected background rates of various
types after event selections are summarized in
Table~\ref{tab:backgroundtable}. No excess of events is observed above
background expectation, both before and after the NR median
selections.  Figure~\ref{fig:dm_band} shows the distribution of
$\log_{10}$(S2/S1) vs. S1 for the DM search data in Run 8 and Run 9.

\begin{table}[h]
\centering
\begin{tabular}{cccc|c|c}
\hline\hline
 & ER & Accidental & Neutron & \parbox[t]{1.4cm}{Total\\Expected} & \parbox[t]{1.4cm}{Total\\observed}\\
\hline
Run 8 & 622.8 & 5.20 & 0.25 & 628$\pm$106 & 734 \\
\hline
\parbox{1.8cm}{Below \\NR median} & 2.0 & 0.33 & 0.09 & 2.4$\pm$0.8 & 2\\
\hline
Run 9 & 377.9 & 14.0 & 0.91 & 393$\pm$46 & 389 \\
\hline
\parbox{1.8cm}{Below \\NR median} & 1.2 & 0.84 & 0.35 & 2.4$\pm$0.7 & 1\\
\hline
\hline
\end{tabular}
\caption{The expected background events in Run 8 and Run 9 in the FV,
before and after the NR median selection.  Number of events from the
data are shown in the last column.}
\label{tab:backgroundtable}
\end{table}

\begin{figure}[h] \centering
  \includegraphics[width=0.6\textwidth]{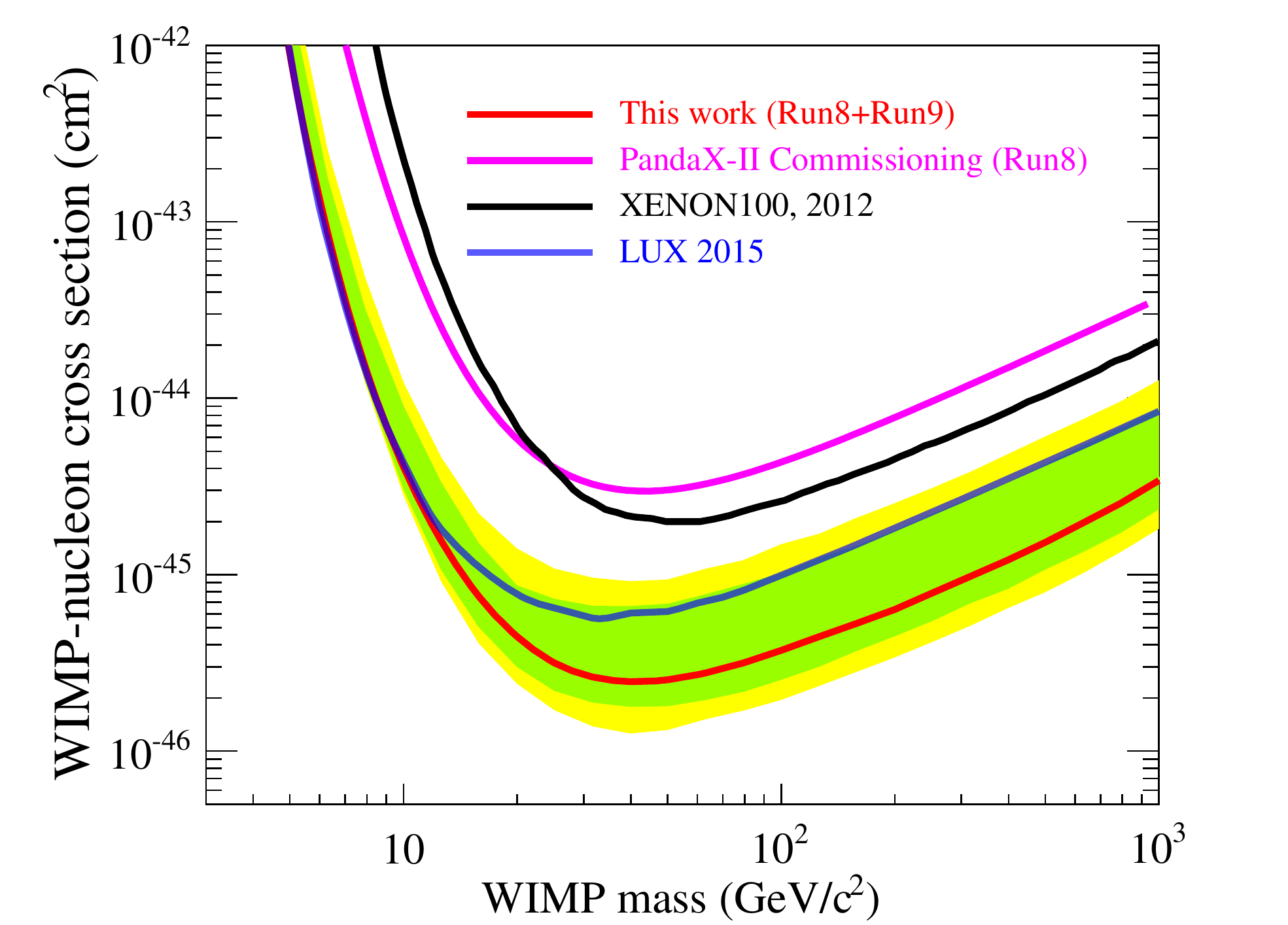}
  \caption{The 90\% CL upper limits for the spin-independent isoscalar
WIMP-nucleon cross sections from the combination of PandaX-II Run 8
and Run 9 (red solid). Selected recent world results are plotted for
comparison: PandaX-II Run 8 results~\cite{Tan:2016diz} (magenta),
XENON100 225 day results~\cite{Aprile:2013doa} (black), and LUX 2015
results~\cite{Akerib:2015rjg}(blue). The 1 and 2-$\sigma$ sensitivity
bands are shown in green and yellow, respectively.}
\label{fig:limit}
\end{figure}

In the absence of possible WIMP signals in data, the 90\% confidence
level (CL) upper limits of the spin-independent isoscalar WIMP-nucleon
cross section are derived and shown in Fig.~\ref{fig:limit}. Our
observed limits lie within the $\pm$1$\sigma$ sensitivity band, and
the lowest limit is 2.5$\times$10$^{-46}$~cm$^2$ at a WIMP mass of 40
GeV/c$^2$, which represents an improvement of more than a factor of 10
from PandaX-I (Ref~\cite{Tan:2016diz}). In the high WIMP mass region,
our results are more than a factor of 2 more stringent than the
previously best results from LUX experiment~\cite{Akerib:2015rjg}.

The PandaX-II collaboration has recently reported the spin-dependent
(SD) WIMP-nucleon upper limits using the same
data~\cite{Fu:2016ega}. Most stringent limits on the WIMP-neutron
cross sections for WIMPs with masses above 10 GeV/c$^{2}$ are set in
all direct detection experiments, with more than a factor of two
improvement on previously best available limits. The PandaX-II
experiment is expected to run till end of 2017. Meanwhile the
collaboration is preparing for a multi-ton scale xenon detector to
further improve the WIMP DM detection sensitivity.

\end{document}